\documentclass[aps,prb,twocolumn,amsmath,amssymb,showpacs]{revtex4}

\usepackage{graphicx}
\usepackage{dcolumn}
\usepackage{bm}

\begin{document}

\title{Variational Monte Carlo study of
ferromagnetism in the two-orbital Hubbard model\\
on a square lattice}

\author{Katsunori Kubo}

\affiliation{
Max Planck Institute for Chemical Physics of Solids,
01187 Dresden, Germany}
\affiliation{
Advanced Science Research Center, Japan Atomic Energy Agency,
Tokai, Ibaraki 319-1195, Japan}

\received{7 November 2008;
revised manuscript received 31 December 2008;
published 27 January 2009}

\begin{abstract}
To understand effects of orbital degeneracy on magnetism,
in particular effects of Hund's rule coupling,
we study the two-orbital Hubbard model
on a square lattice by a variational Monte Carlo method.
As a variational wave function, we consider a Gutzwiller projected
wave function for a staggered spin and/or orbital ordered state.
We find a ferromagnetic phase with staggered orbital order around quarter-filling,
i.e., electron number $n=1$ per site,
and an antiferromagnetic phase without orbital order around half-filling $n=2$.
In addition, we find that
another ferromagnetic phase without orbital order
realizes in a wide filling region for large Hund's rule coupling.
These two ferromagnetic states are metallic except for quarter filling.
We show that orbital degeneracy and strong correlation effects
stabilize the ferromagnetic states.

\end{abstract}



\maketitle

Mechanism of itinerant ferromagnetism is a long standing problem
in physics of condensed matter.
As a simple model for itinerant ferromagnetism,
the single-orbital Hubbard model has been studied intensively,
but it has been revealed that it is
difficult to stabilize a ferromagnetic state
in the Hubbard model with only nearest-neighbor hopping
on simple lattices such as a square lattice.

One possible improvement to the Hubbard model for ferromagnetism
is a modification of the band structure.
Since the early stage of the study of ferromagnetism,
a large density of states around Fermi level,
as in an fcc lattice with appropriate filling,
has been suggested to stabilize a ferromagnetic state.~\cite{Gutzwiller,Hubbard,Kanamori}
Indeed, realization of ferromagnetic ground states is proven
for some flat band systems~\cite{Mielke1,Mielke2}
and nearly flat band systems.~\cite{Tasaki}
It is also shown that
ferromagnetism occurs for an fcc-type infinite dimensional lattice
and for an fcc lattice by using a dynamical mean-field theory,~\cite{Ulmke}
while a dynamical mean-field theory for a hypercubic lattice
does not show ferromagnetism.~\cite{Freericks}
For finite dimensions, it has been shown that
ferromagnetism can occur by including next-nearest hopping,
which induces Van Hove singularity,
for a chain~\cite{Daul} and for a square lattice.~\cite{Hlubina}

Another possible improvement is inclusion of orbital degree of freedom,
which may be important to deal with realistic situations in transition metals.
For orbitally degenerate systems,
it has been suggested that intra-atomic Hund's rule coupling
can stabilize ferromagnetism.~\cite{Slater,Zener,Roth}
%
The simplest extended model of the single-orbital Hubbard model
including orbital degree of freedom
is the two-orbital Hubbard model.
This model shows ferromagnetism with antiferro-orbital order
at quarter-filling, i.e., electron number $n=1$ per site,
in the strong Coulomb interaction limit.~\cite{Kugel}

This ferromagnetic state at $n=1$ is insulating.
Thus, it is an interesting problem as to what extent is the ferromagnetic state stable
against doping of electron or hole which makes the system metallic.
It is found that the ferromagnetic state is stable to some extent against doping
in one dimension~\cite{Sakamoto}
and in infinite dimensions.~\cite{Held,Momoi}
For other finite dimensions, there are few studies on
doping effects on magnetism of the two-orbital Hubbard model.
Sakai \textit{et al.}~\cite{Sakai} have studied the two-orbital Hubbard model
on an fcc lattice by a dynamical mean-field theory,
and have stressed importance of the lattice structure and Hund's rule coupling
for ferromagnetism.

To understand magnetism of the two-orbital Hubbard model deeply,
we have to investigate the model for many parameter sets,
since the two-orbital Hubbard model has a parameter for Hund's rule coupling
in addition to that for the Coulomb interaction.
However, such an extensive study has been difficult for a two-orbital model
beyond the Hartree-Fock approximation,
since there are as many as sixteen electron configurations at each site
and it is hard for numerical calculations.

In this paper, to overcome such difficulty,
we apply a variational Monte Carlo method~\cite{Yokoyama}
to the two-orbital Hubbard model on a square lattice.
We use a Gutzwiller projected wave function as a variational wave function.
This wave function is simple enough but includes correlation effects,
and we can evaluate energy for several states and for various parameters.
In particular, we can construct a phase diagram by varying
the value of Hund's rule coupling and filling $n$.
Thus, we can investigate overall feature of the two-orbital Hubbard model.
At $n=1$, a similar model without considering orbital order~\cite{Kobayashi1}
and the two-orbital Hubbard model considering possibility of orbital order~\cite{Kubo}
have been studied by the variational Monte Carlo method,
but the effect of doping has not been investigated by these studies,
which is a main topic of the present paper.

The two-orbital Hubbard model is given by
\begin{equation}
  \begin{split}
    H=&\sum_{\mathbf{k},\tau,\sigma}
    \epsilon_{\mathbf{k}}
    c^{\dagger}_{\mathbf{k} \tau \sigma}c_{\mathbf{k} \tau \sigma}
    +U \sum_{i, \tau}
    n_{i \tau \uparrow} n_{i \tau \downarrow}\\
    &+U^{\prime} \sum_{i}
    n_{i 1} n_{i 2}
    + J \sum_{i,\sigma,\sigma^{\prime}}
    c^{\dagger}_{i 1 \sigma}
    c^{\dagger}_{i 2 \sigma^{\prime}}
    c_{i 1 \sigma^{\prime}}
    c_{i 2 \sigma}
    \\
    &+ J^{\prime}\sum_{i,\tau \ne \tau^{\prime}}
    c^{\dagger}_{i \tau \uparrow}
    c^{\dagger}_{i \tau \downarrow}
    c_{i \tau^{\prime} \downarrow}
    c_{i \tau^{\prime} \uparrow},
  \end{split}
  \label{eq:H}
\end{equation}
where $c_{i\tau\sigma}$ is the annihilation operator of
the electron at site $i$ with orbital $\tau$ ($=1$ or 2)
and spin $\sigma$ ($=\uparrow$ or $\downarrow$),
$c_{\mathbf{k}\tau\sigma}$ is the Fourier transform of it,
$n_{i \tau \sigma}=c^{\dagger}_{i \tau \sigma} c_{i \tau \sigma}$, and
$n_{i \tau}=\sum_{\sigma}n_{i \tau \sigma}$.
The coupling constants $U$, $U^{\prime}$, $J$, and $J^{\prime}$
denote the intra-orbital Coulomb, inter-orbital Coulomb, exchange,
and pair-hopping interactions, respectively.
We use the relation
$U=U^{\prime}+J+J^{\prime}$,
which is satisfied in several orbital-degenerate models
such as a model for $p$-orbitals, a model for $e_g$ orbitals,
and a model for $t_{2g}$ orbitals.~\cite{Tang}
We also use the relation
$J=J^{\prime}$,
which holds if we can choose wave functions of orbitals real.~\cite{Tang}
We consider only a nearest-neighbor hopping
integral $t$ for both orbitals,
and the kinetic energy is given by
$\epsilon_{\mathbf{k}}=2t(\cos k_x+\cos k_y)$.
Here we have set the lattice constant unity.

We consider the variational wave function given by
\begin{equation}
  | \Psi \rangle = P_{\text{G}} | \Phi \rangle
  =\prod_{i \gamma}
  [1-(1-g_{\gamma})|i\gamma\rangle \langle i\gamma |] | \Phi \rangle,
\end{equation}
where $P_{\text{G}}$ is the Gutzwiller projection operator
for onsite density correlation.~\cite{Okabe,Bunemann,Kobayashi2}
$|i\gamma\rangle \langle i\gamma|$ denotes projection onto the state $\gamma$
at site $i$ and $g_{\gamma}$ is the variational parameter
controlling the probability of state $\gamma$.
There are sixteen states at each site in the present two-orbital model.
The Hartree-Fock type wave function $| \Phi \rangle$,
which describes a charge, spin, orbital, and spin-orbital coupled
ordered state, is given by
\begin{equation}
  |\Phi \rangle = \prod_{\mathbf{k} a \tau \sigma}
  b^{(a) \dagger}_{\mathbf{k} \tau \sigma} |0 \rangle,
\end{equation}
where $| 0 \rangle$ is the vacuum.
The quasiparticles occupy $N_{\sigma}$ states for each spin $\sigma$
from the lowest quasiparticle energy state,
where $N_{\sigma}$ is the number of electrons with spin $\sigma$.
The energy of the quasiparticle in the ordered state is given by
\begin{equation}
  \lambda^{(a)}_{\mathbf{k} \tau \sigma}=a\sqrt{\Delta^2_{\tau \sigma}+\epsilon^2_{\mathbf{k}}}.
\end{equation}
The creation operators of quasiparticles are given by
\begin{align}
  b^{(-) \dagger}_{\mathbf{k} \tau \sigma}
  &=u_{\mathbf{k} \tau \sigma} c^{\dagger}_{\mathbf{k} \tau \sigma}
  +\text{sgn}(\Delta_{\tau \sigma})v_{\mathbf{k} \tau \sigma} c^{\dagger}_{\mathbf{k}+\mathbf{Q} \tau \sigma},\\
  b^{(+) \dagger}_{\mathbf{k} \tau \sigma}
  &=-\text{sgn}(\Delta_{\tau \sigma})v_{\mathbf{k} \tau \sigma} c^{\dagger}_{\mathbf{k} \tau \sigma}
  +u_{\mathbf{k} \tau \sigma} c^{\dagger}_{\mathbf{k}+\mathbf{Q} \tau \sigma},
\end{align}
where $\mathbf{Q}=(\pi,\pi)$ is the ordering vector considered in this study and
\begin{align}
  u_{\mathbf{k} \tau \sigma}=\left[ \left( 1-\epsilon_{\mathbf{k}}/\sqrt{\Delta^2_{\tau \sigma}+\epsilon^2_{\mathbf{k}}} \right) /2 \right]^{1/2},\\
  v_{\mathbf{k} \tau \sigma}=\left[ \left( 1+\epsilon_{\mathbf{k}}/\sqrt{\Delta^2_{\tau \sigma}+\epsilon^2_{\mathbf{k}}} \right) /2 \right]^{1/2}.
\end{align}
The quasiparticle gap in the ordered state is given by
\begin{equation}
  \begin{split}
    \Delta_{\tau \sigma}=&\Delta_{\text{c}}+\Delta_{\text{s}}(\delta_{\sigma \uparrow}-\delta_{\sigma \downarrow})
    +\Delta_{\text{o}}(\delta_{\tau 1}-\delta_{\tau 2})\\
    &+\Delta_{\text{so}}(\delta_{\sigma \uparrow}-\delta_{\sigma \downarrow})(\delta_{\tau 1}-\delta_{\tau 2}),
  \end{split}
  \label{eq:gap}
\end{equation}
where $\Delta_{\text{c}}$, $\Delta_{\text{s}}$, $\Delta_{\text{o}}$, and $\Delta_{\text{so}}$
denote the gaps for charge, spin, orbital, and spin-orbital ordered states, respectively,
and we take them as variational parameters.

Here, we have chosen $z$-component of spin for the ordered state.
We can choose $x$- or $y$-component,
but they are equivalent due to the rotational symmetry in the spin space.
On the other hand, there is rotational symmetry only in the $z$-$x$ plane
in the orbital space.
For orbital order,
in addition to $z$-component as in Eq.~\eqref{eq:gap},
we have also investigated possibility of $y$-component order.
The model Hamiltonian~\eqref{eq:H} can be rewritten
in terms of basis states of $y$-component of orbital
by replacing interaction parameters with tilde:
$\tilde{U}        =( U+U^{\prime}+J-J^{\prime})/2$,
$\tilde{U}^{\prime}=( U+U^{\prime}-J+J^{\prime})/2$,
$\tilde{J}        =( U-U^{\prime}+J+J^{\prime})/2$, and
$\tilde{J}^{\prime}=(-U+U^{\prime}+J+J^{\prime})/2$.
Thus we can study $y$-component orbital order
with the same form for the variational function
by simply changing interaction parameters.

We evaluate the expectation value of energy for the variational wave function
by using the Monte Carlo method,
and optimize these gap parameters and the Gutzwiller parameters to minimize energy.
For the optimization we use a fixed sampling method.~\cite{Umrigar,Giamarchi}
The number of parameters in the Gutzwiller projection
can be reduced from sixteen to ten by considering
equivalence of the two orbitals.~\cite{Kobayashi2}
We can further reduce the number of the Gutzwiller parameters to seven
when we consider conservation of spin, i.e., when we fix $N_{\sigma}$.
Thus the total number of the variational parameters is eleven.
Note that, in this study, we take partial ferromagnetic states into consideration
and spin $\uparrow$ and $\downarrow$ are not equivalent for these states.
Thus we cannot reduce the number of the parameters further
by considering spin states.
We can also evaluate energy by fixing some parameters, for example,
we set all the gap parameters zero for a paramagnetic state.
The calculations have been done for a $12 \times 12$ lattice
with periodic-antiperiodic boundary conditions.

\begin{figure}[t]
  \includegraphics[width=\linewidth]{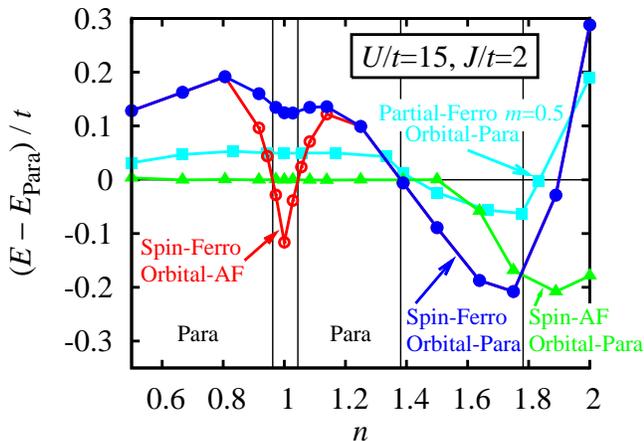}
  \caption{\label{figure:n_dep}
    (Color online)
    Filling dependence of energy for several states
    measured from that of the paramagnetic state:
    Spin-AF Orbital-Para ($m=0$, $\Delta_{\text{s}}\ne 0$, $\Delta_{\text{o}}=0$, solid triangles),
    Partial-Ferro $m=0.5$ Orbital-Para ($\Delta_{\text{s}}=0$, $\Delta_{\text{o}}=0$, solid squares),
    Spin-Ferro Orbital-Para ($m=1$, $\Delta_{\text{s}}=0$, $\Delta_{\text{o}}=0$, solid circles),
    and
    Spin-Ferro Orbital-AF ($m=1$, $\Delta_{\text{s}}=0$, $\Delta_{\text{o}}\ne 0$, open circles).
  }
\end{figure}

Figure~\ref{figure:n_dep} shows energy $E$ in ordered states per site
measured from energy $E_{\text{Para}}$ per site in the paramagnetic state
as functions of $n$ for $U/t=15$ and $J/t=2$ as an example.
Statistical errors are much smaller than the symbol sizes.
In the low filling region, the paramagnetic state is stable.
Around quarter-filling $n=1$, the ferromagnetic phase with
antiferro-orbital order appears.
In the filling region $1.38 \lesssim n \lesssim 1.78$,
the ferromagnetic phase without orbital order appears.
Around half-filling $n=2$, antiferromagnetic phase appears
due to nesting of the Fermi surface.
While we have calculated energy of partial ferromagnetic states,
for example, for $m=(N_{\uparrow}-N_{\downarrow})/(N_{\uparrow}+N_{\downarrow})=0.5$
shown in Fig.~\ref{figure:n_dep},
these states do not become the ground state.

Here we comment on the ground states at $n=1$ and 2.
The chemical potential $\mu$ can be obtained from $\mu=dE/dn$
in the ground state and we also obtain the relation
\begin{equation}
  \frac{dn}{d\mu} \frac{d^2E}{dn^2}=1.
\end{equation}
Thus, in an insulating state, i.e., $dn/d\mu=0$,
the second derivative of energy $E$ with respect to $n$
should diverge and vice versa.
At $n=1$, the ground state energy has a cusp as shown in Fig.~\ref{figure:n_dep}
and the ground state is insulating.
Note that we can obtain energy for $n>2$ from the present data
by using electron-hole symmetry of the model,
and we find a cusp in the ground state energy also at $n=2$.

\begin{figure}[t]
  \includegraphics[width=\linewidth]{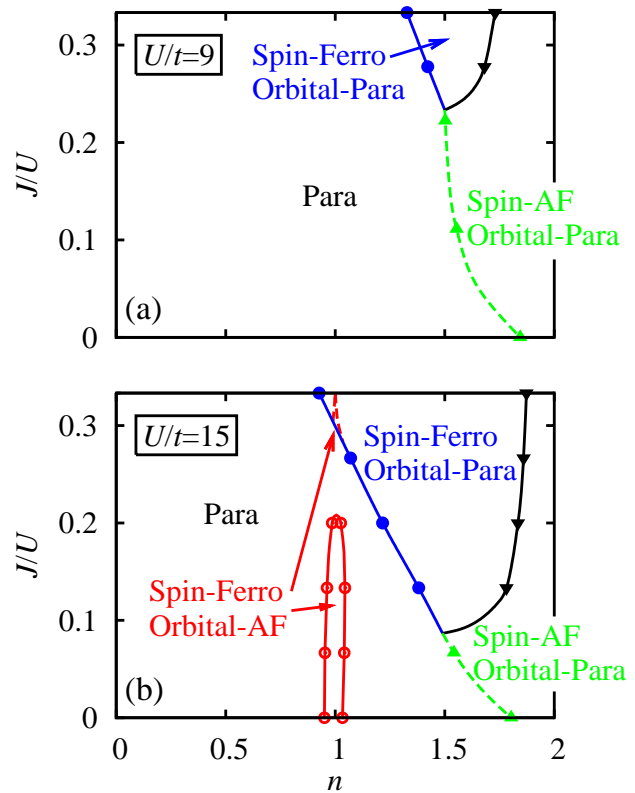}
  \caption{\label{figure:PD}
    (Color online)
    Phase diagrams for (a) $U/t=9$ and for (b) $U/t=15$.
    Solid lines denote first-order transitions
    and
    dashed lines denote second-order transitions.
  }
\end{figure}

Figure~\ref{figure:PD} shows phase diagrams for $U/t=9$ and 15.
First, we explain the phase diagram Fig.~\ref{figure:PD}(b) for $U/t=15$.
The ferromagnetic phase with antiferro-orbital order
appears around quarter-filling as is expected from
the effective Hamiltonian for the strong coupling limit.
However, it is found that this ferromagnetic phase is easily
destabilized by doping.
To stabilize this ferromagnetic phase in a wider filling region,
much larger value of Coulomb interaction is necessary.

This ferromagnetic phase is also destabilized
by increasing Hund's rule coupling $J$.
Hund's rule coupling is usually expected to stabilize magnetically ordered states, 
but the effective interaction $U_{\text{eff}}=U^{\prime}-J$
between different orbitals is reduced by Hund's rule coupling.
Thus, Hund's rule coupling destabilizes the orbital order,
and as a result, the ferromagnetic state supported by the orbital order
is also destabilized.
Note that in the ferromagnetic phase,
there is rotational symmetry in the orbital space,
and orbital order for $x$-, $y$-, and $z$-components are equivalent.

At higher filling region, another ferromagnetic phase without orbital order appears
in a large Hund's rule coupling region.
The ferromagnetic phase extends in a wide parameter region for $n \gtrsim 1$,
while not for $n \lesssim 1$.
This finding is in agreement with the statement that
double-exchange type mechanism works well for $n \gtrsim 1$
since the probability of double occupancy is high,
but it is less effective for $n \lesssim 1$.~\cite{Momoi}

In this ferromagnetic phase at $n \simeq 1$ and $J/U \simeq 0.3$,
we expect orbital-antiferro order,
since in the ferromagnetic state the model is reduced
to the single-orbital Hubbard model with effective interaction $U_{\text{eff}}$
if we regard spin in the single-orbital Hubbard model as orbital
and around $n=1$ an orbital-antiferro state should occur.
However, it is difficult to distinguish a small energy difference
between orbital-para and orbital-antiferro states
around there due to a small value of $U_{\text{eff}}$.
Thus, the phase boundary between spin-ferro orbital-para
and spin-ferro orbital-antiferro in Fig.~\ref{figure:PD}(b)
is merely a eye guide.

Around half-filling, the antiferromagnetic phase appears
as is expected from the nesting of Fermi surface.
The phase transition from the paramagnetic phase to the antiferromagnetic phase
is second-order.
We have checked that the energy difference between
these phases is proportional to $(n-n_{\text{c}})^2$ for $n \gtrsim n_{\text{c}}$,
where $n_{\text{c}}$ is the critical filling.

Note that the spin-ferro orbital-antiferro state at $n=1$
and the spin-antiferro orbital-para state at $n=2$ are insulating,
and other ground states are metallic.

By reducing the Coulomb interaction $U$,
the regions of the ordered phases become narrower
as shown in Fig.~\ref{figure:PD}(a) for $U/t=9$.
In particular the spin-ferro orbital-antiferro state disappeared.
At $n=1$, the ferromagnetic state with orbital order disappears at $U/t \lesssim 10$.~\cite{Kubo}
The other ferromagnetic phase without orbital order is also easily destabilized by
reducing the Coulomb interaction.
This fact indicates that realization of ferromagnetism is a strong correlation effect.
On the other hand,
the antiferromagnetic phase around half-filling, which is stabilized by
the nesting of the Fermi surface and can be obtained with a weak-coupling theory,
realizes in a wide region even at $U/t=9$.

Note that
we have also calculated energy of the single-orbital Hubbard model
within the Gutzwiller wave function,
and we have found that
a much larger value of Coulomb interaction $U/t \gtrsim 23$
is necessary to stabilize a ferromagnetic phase.
Thus, the orbital degeneracy and Hund's rule coupling are
important ingredients for realization of ferromagnetism
with a moderate value of the Coulomb interaction.

To summarize,
we have studied the two-orbital Hubbard model on a square lattice
by a variational Monte Carlo method.
We have considered charge, spin, orbital, and spin-orbital coupled
ordered states for the variational wave function.
Then, we have constructed phase diagrams for the ground states.
We find a narrow region of the ferromagnetic state with orbital-antiferro order
around quarter-filling
and a wide region of ferromagnetic phase without orbital order
at large Hund's rule coupling
for $U/t=15$.
The ferromagnetic phase with orbital order is easily destabilized
by doping and by reducing the Coulomb interaction.
The ferromagnetic phase without orbital order is also destabilized
strongly by reducing the Coulomb interaction.
Thus, realization of ferromagnetic states is a strong correlation effect.
Investigation of effects of realistic anisotropic hopping integral
depending on orbital
and further improvement of the variational wave function
are important future problems.

The author thanks T. Takimoto and P. Thalmeier
for reading the manuscript and useful comments.
This work is supported from Japan Society for the Promotion of Science
through a Postdoctoral Fellowship for Research Abroad.


\end{document}